\documentstyle[emulateapj,onecolfloat,psfig]{article}

\newlength{\tskip}
\newlength{\colwidth}

\newcommand{\beq}{\begin{equation}}
\newcommand{\eeq}{\end{equation}}
\newcommand{\beqa}{\begin{eqnarray}}
\newcommand{\eeqa}{\end{eqnarray}}

\newcommand{\LISA}{\emph{LISA~}}

\begin{document}
\twocolumn[
\title{Optical Identification of Close White Dwarf Binaries in the LISA Era}
\author{Asantha Cooray, Alison J. Farmer and Naoki Seto}
\affil{Theoretical Astrophysics, Division of Physics, Mathematics and Astronomy, 
California Institute of Technology, MS 130-33, Pasadena, CA 91125\\}

\begin{abstract}
The \emph{Laser Interferometer Space Antenna} (\emph{LISA}) is expected to
 detect close white dwarf binaries (CWDBs) through their gravitational
 radiation. Around 3000 binaries will be spectrally resolved at
 frequencies $> 3$ mHz, and their positions on the sky will be
 determined to an accuracy ranging from a few tens of arcminutes to a
 degree or more.  Due to the small
 binary separation, the optical light curves of $\gtrsim$ 30\% of
 these  CWDBs are expected to show eclipses, giving a unique signature
 for identification in follow-up studies of the \LISA error boxes.  While
 the precise optical location improves binary parameter
 determination with \LISA data, the optical light curve captures additional physics of
 the binary, including the individual sizes of the stars in terms of the
 orbital separation.  To optically identify a substantial fraction of CWDBs and
thus localize them very accurately, a rapid monitoring campaign is required, capable of
 imaging a square degree or more in a reasonable time, at intervals of
 10--100 seconds, to magnitudes between 20 and 25.
 While the detectable fraction can be up to many tens of percent of the total
 resolved \LISA CWDBs, the exact fraction is uncertain due to unknowns
 related to the white dwarf spatial distribution, and potentially
 interesting physics, such as induced tidal heating of the WDs due to
 their small orbital separation.
\end{abstract}

 ]

\section{Introduction}
The \emph{Laser Interferometer Space Antenna} (\emph{LISA}) is expected to
establish the presence of low frequency gravitational wave (GW)
sources. \emph{LISA}'s frequency coverage between 10$^{-1}$ and
10$^{-4}$ Hz is ideal for the detection of GWs from close white dwarf
binaries (CWDBs; e.g., Hils, Bender \& Webbink 1990; Farmer \& Phinney
2003; Nelemans et al. 2001b).
While the study of the closest white dwarf binaries currently implies
theoretical prediction, in the \LISA era direct observations will
provide a wealth of data that are expected to greatly expand our
knowledge of the Galactic CWDB population and its
evolutionary history.

The study of CWDBs using \LISA GW data has been considered
previously. At frequencies above $f_{\rm{res}}\sim 3$ mHz, most CWDBs
will be spectrally resolved (Cornish \& Larson 2002),  and one expects
a detection of $\sim$ 3000 $(f_{\rm{res}}/3 \; {\rm mHz})^{-8/3}$
binaries (e.g., Hils, Bender \& Webbink 1990; Seto 2002 and references
therein) above this frequency. Below $f_{\rm{res}}$ there will be
other individually resolved binaries, particularly those that are
nearby, but here we focus on those above the resolution limit, as
theoretical 
predictions of the resolved number are more reliable and parameter
extractions are likely to be more useful for localizing the CWDB
population.
The accuracy to which we can determine the position on the sky of a
given binary depends on the level at which its emitted GW are detected
above the noise (e.g., Cutler 1998). In addition to the location, GW
allow one to establish certain parameters related to the binary, such
as the chirp mass, distance, period, and the orientation of the binary
with respect to the observer (Takahashi \& Seto 2002).



It is clearly advantageous to combine the GW data with any additional
data available, to reduce the number of parameters extracted from GW
data alone and thus improve their accuracy.  For example, if
electromagnetic radiation emitted by the binary can be used to
localize it, then we expect improvements to the extraction of the
other parameters.  Following Takahashi \& Seto (2002), if the exact
location of the binary on the sky is known a priori, then the accuracy
of most parameters extracted from the first year of \LISA data
improves roughly by a factor of 2 to 3, while there is more than an
order of magnitude improvement in the determination of the direction
of the binary's angular momentum vector.

 An obvious way to localize CWDBs involves follow-up optical
observations of the \LISA error box: once the location is established
to the precision of the optical resolution, one can
revisit the \LISA data to extract improved binary parameters. 
The \LISA error box, however, can be large and traditional techniques for
identifying the faint candidate white dwarfs in such a large area will
be cumbersome. They will also likely be inconclusive, as there will be
many such candidates in a large error box. Fortunately, due to the
small binary separation, many of these binaries will be eclipsing,
giving rise to unique signatures in their light curves. Each eclipsing
system will display two transits per period, i.e. periodic dimming
will occur at the GW frequency. Since WD radii are not strongly
dependent on their masses, a centrally eclipsing binary ought to
produce at least one transit per period of depth more than a few tens
of percent. Because the period and orbital phase are already
determined from \LISA data, a clear identification can be made.


In addition to the precise position, optical observations of CWDBs
provide additional information on the binary that cannot be extracted
from gravitational wave data alone, including the two stellar radii,
in terms of the orbital separation and the binary orientation.
While there is a strong motivation for
optical follow-up observations, it is as yet unclear to what extent
this can be achieved and what will be required in terms of an
observing program. The purpose of this {\it Letter} is to discuss
these issues in detail. The discussion is organized as follows: In the next section,
we briefly discuss the properties of CWDBs that will be detectable
with \LISA and then describe the optical light curves associated with
binary eclipses. We then discuss the requirements for an optical
follow-up campaign and conclude with a summary.

\begin{figure}[t]
\psfig{file=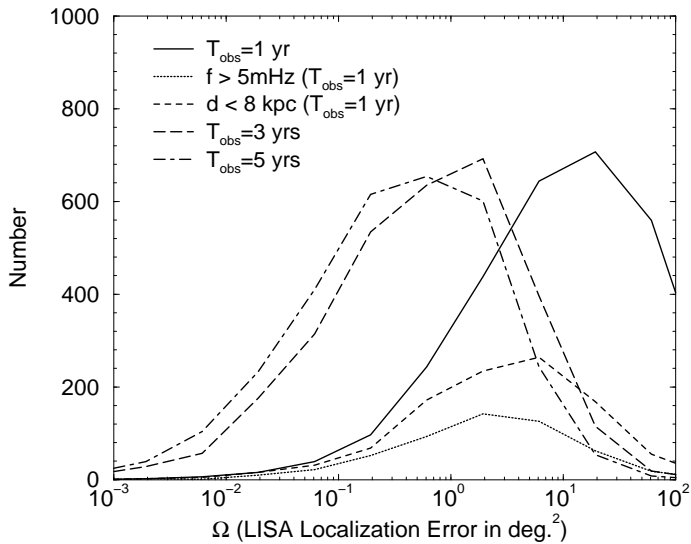,width=3.6in,angle=-90}
\caption{The LISA localization error distribution for 3000 Galactic
 CWDBs with $f>3$mHz. We define the angular localization error $\Omega$ such that the
probability of a source lying outside an error ellipse $\Delta Omega$ becomes $\exp(-\Delta \Omega/\Omega)$ (Cutler 1998).
These errors are calculated following the Fisher
 matrix analysis of  Takahashi \& Seto (2002) and assuming for an
 observation time of 1 (solid line), 3  (dashed line) and 5 (dot-dashed line) years with
 LISA. Using just the first year of data, the localization errors are
 at the level of a few sqr. degrees; this improves by more than a
 factor of $\sim$ 10 when 3 years of data are used. The improvement from
 3 to 5 years is minor when compared to the improvement from 1 year to 3 years. For comparison,
we also show the localization error for subsamples with frequencies above 5 mHz and at distances
below 8 kpc.}
\label{fig:area}
\end{figure}

\begin{figure}[t]
\psfig{file=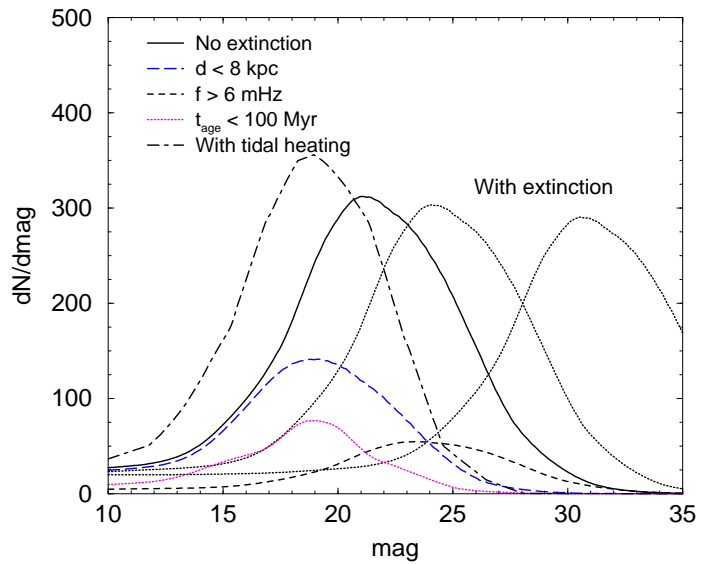,width=3.6in,angle=-90}
\caption{The expected V-band magnitude distribution of the brightest
member of the close white dwarf binaries resolved with LISA (solid
line). This distribution must be corrected for Galactic extinction and
we consider a simple prescription following Bahcall \& Soneira (1980);
in decreasing brightness, the dashed lines show the magnitude
distribution when CWDBs are assumed to be distributed uniformly in a
thick disk with scale height $\sim$ 1.5 kpc, and the case in which the distribution
is highly flattened, with scale height $\sim$ 200 pc, such that more are towards the galactic plane.  In
the latter case, the extinction is significant and presents a
challenge for optical studies. It is more likely that the true
distribution is somewhere between the two cases we have considered. We
also expect a small fraction of CWDBs to be distributed in the halo
and not be affected by extinction significantly.  With a dot-dashed
line, we show the magnitude distribution (with no extinction) when we
include an increase in white dwarf luminosity by tidal heating; we
have considered the maximum brightening here under the assumption of
equal-mass binaries and following the approach in Iben et
al. (1998). The long-dashed line shows the distribution of magnitudes
for white dwarfs at distances less than 8 kpc, while the dotted lines
show the magnitude distribution of white dwarf binaries with GW
frequencies greater than $\sim$ 6 mHz.}
\label{fig:spec}
\end{figure}

\section{\emph{LISA} Measurements and Optical Follow-up}

First, we will review CWDB detection with \LISA and then move on to
discuss aspects related to their optical light curves.  Briefly, one
observes the two GW components given by the quadrupole approximation
in the principal polarization coordinate (Peters 1964)
\begin{eqnarray}
h_{+}(t) &=& A \cos \left[ 2\pi \left(f+\frac{1}{2} \dot{f} t \right)
t + \phi(t) \right] \times \left[1+ \cos^2 i\right] \nonumber \\
h_{\times}(t) &=& -2A \sin \left[ 2\pi \left(f+\frac{1}{2} \dot{f} t
\right)t + \phi(t)\right] \times \cos i \nonumber \\
\end{eqnarray}
where $\cos i$ is the inclination angle of the binary with respect to
the observer and $\phi(t)$ is the phase resulting from the Doppler
phase due to the revolution of LISA around the Sun and an integral constant ($\phi_0$).

CWDBs are expected to have circular orbits due to tidal interaction in
earlier evolutionary stages.  The amplitude of the wave is given by
\begin{equation}
A = \frac{5}{96 \pi^2} \frac{\dot{f}}{f^3 D} \, ,
\label{amplitude}
\end{equation}
where $D$ is the distance to the source.  Note that the GW frequency
($=2/P_{\rm orb}$ where $P_{\rm orb}$ is the orbital period) for a
circular orbit is related to the total mass $M_1+M_2$ and the
separation $a$  of the binary via
\begin{equation}
f  \approx  3.5 \times 10^{-3} \left(\frac{M_{\rm tot}}{0.9
M_{\sun}}\right)^{1/2} \left(\frac{a}{10^5 {\rm km}}\right)^{-3/2} \;
{\rm Hz}\, ,
\label{eqn:freq}
\end{equation}
while the time variation of this frequency is
\begin{eqnarray}
&&\dot{f} = \frac{96 \pi^{8/3}}{5} {\tilde M}^{5/3}  f^{11/3}
\nonumber \\ &\approx& 1.2 \times 10^{-16} \left(\frac{\tilde M}{0.4
M_{\sun}}\right)^{5/3} \left(\frac{M_{\rm tot}}{0.9
M_{\sun}}\right)^{11/6} \left(\frac{a}{10^5 {\rm km}}\right)^{-11/2}
{\rm Hz}\; {\rm s}^{-1} \, ,
\label{chirp}
\end{eqnarray}
when the chirp mass is given by ${\tilde M} = M_1^{3/5} M_2^{3/5} /
(M_1+M_2)^{1/5}$ and the total mass $M_{\rm tot}=M_1+M_2$.

With GW data, one can estimate a total of 8 separate quantities, $A$, $f$, $\dot{f}$,
$\phi_0$, location ($\vec n$, 2 parameters), and the direction of the
angular momentum, $\vec L$ (2 parameters), though these estimates are not independent.
The orbital
inclination angle is given by $\cos^{-1} (\vec n \cdot \vec L)$.  In
terms of physical quantities of interest, with $A$, $f$, $\dot{f}$,
one extracts $D$ and ${\tilde M}$; in addition to these parameters one
also constraints a combination of $M_{\rm tot}$ and $a$ using the
frequency information following Eq.~\ref{eqn:freq}. Note that the
relations (\ref{amplitude}) and (\ref{chirp}) are given for Newtonian
point particles as corrections for the finiteness of the stars are not
significant. 

The localization of binaries with \LISA observations is discussed in
Cutler (1998) for monochromatic sources (with $\dot{f}=0$), and in
Takahashi \& Seto (2002) for general observational situations.
Following the Fisher matrix calculation in the latter paper, in
Fig.~1, we show the distribution of error boxes for the binary
locations extracted from the \LISA data, for observations of 1, 3 and 5
years' duration. To produce Fig.~1, we performed a Monte-Carlo
analysis for 3000 Galactic CWDBs with $f>3$mHz, distributed in the
Galaxy according to
\begin{equation}
\rho(R,z) = \rho_0 {\rm sech}\left(\frac{|z|}{z_0}\right)^2
\exp\left(-\frac{R}{ R_0}\right) \, ,
\label{dist}
\end{equation}
with $z_0 = 200$ pc and $R_0=2.5$ kpc (as in Nelemans et al. 2001).
 The \LISA noise curve
is taken from Finn \& Thorne (2000) and the chirp mass distribution,
as a function of GW frequency, is generated for the Galaxy using the
same numerical codes used in Farmer \& Phinney (2003), with a constant
star formation rate over the past 10 Gyr, and other parameters as for
their preferred Model A. 

The locations for Fig.~1 were determined concurrently with the six
other binary parameters. The parameter extraction can be considerably
improved if a precise location for the binary is a priori known, especially if the
observational duration is less than two years.  This
is certainly the case, for example, for the orientation of the orbit,
which has strong a correlation with the location determination.  A
localization would traditionally come from optical observations and we
consider this possibility here. While it may also be possible to
localize CWDBs at other wavelengths of electromagnetic radiation, such
as X-rays, we do not consider this possibility in the present
discussion, but encourage others to do so.

Eclipses in the light curve as each white dwarf transits across the
other will be a unique identifying signature of the binary. There will
be two transits per orbital period. The WDs may or may not have
similar surface temperatures, with the transits in which the second WD
formed recently unlikely to have similar depths. Observationally
WD--WD binaries are seen to be clustered around a mass ratio of unity (Maxted et al. 2002),
so their radii are similar. Even if they are not, the worst-case
scenario for central transit depth is of order a few tens of percent,
for a large mass ratio and equal temperature (or large temperature
ratio) system.

 For a given binary, one can write down the probability for eclipse
along our line of sight as
\begin{eqnarray}
{\rm Prob} &=& \frac{\int_0^{\cos i_{\rm min}} d(\cos i)}{\int_0^1
d(\cos i)} = \cos i_{\rm min} \equiv \frac{R_1+R_2}{a} \nonumber \\
&\sim& 0.30 \left(\frac{f}{3.5 {\rm mHz}}\right)^{2/3}
\left(\frac{M_{\rm tot}}{0.5 M_{\sun}}\right)^{-1/3}
\left(\frac{R_{\rm WD}}{10^4 {\rm km}}\right) \, , \nonumber \\
\label{eqn:prob}
\end{eqnarray}
where $R_1$ and $R_2$ are the WD radii and $a$ is the binary
  separation. The probability is substantial because $a$ is at most
  $\sim 5 (R_1+R_2)$ for spectrally resolved \LISA binaries.  This
  suggests that eclipses will be expected in the light curves of few
  tens of percent of all CWDBs resolved and localized with LISA at
  $f>3$mHz. We note however that this includes shallow grazing
  eclipses, and the probability for near-central transits is around
  half of the above value.  The numerical estimate on the second line
  of Eq.~\ref{eqn:prob} assumes equal mass, and thus equal size WDs of
  radii $R_{\rm WD}$. 

Ignoring subtle effects such as the brightness distribution on the
stellar surfaces, the optical light curves with eclipses, in general,
allow one to establish the following parameters:  $R_1/a \cos i$,
$R_2/a \cos i$, and the ratio, $R_1/R_2$, if the transit durations can
be measured.  Unlike the GW data, which constrain the component masses
through e.g. the chirp mass, transit light curves provide constraints
on the objects' physical sizes.

While there is a high probability of eclipse along our line of sight,
we must consider what it will take to observe them. For this purpose,
we first estimate the V-band apparent magnitude distribution of the brightest
member of the CWDBs following the same mass and age distributions that
were used for the calculation in Fig.~1.  We summarize our results in
Fig. 2.  The luminosities are calculated based on age distributions,
as a function of mass, by taking in to account WD cooling based on
numerical fits to results of Bl\"ocker (1995) and Driebe et al. (1998) in
Nelemans et al. (2001a). We take a uniform bolometric correction of
-0.2 consistent with DA white dwarfs at 8000K  with $\log g=8$, consistent with the age
distribution (Rohrmann 2001); for a range of temperatures, the corrections 
are at the level of -0.1 to -0.5 and we ignore this distribution for simplicity given the large
uncertainties in the extinction as discussed below. 
The distribution is normalized to a total of 3000 CWDBs
that are expected to be localized with \LISA at $f>3$mHz (Seto
2002). This number, however, is uncertain and can be up to $\sim$ 6000
in different models (e.g., Nelemans et al. 2001b). The number depends
also on the observational period.  To convert
luminosities to apparent magnitudes, we distribute these binaries in
our galaxy as in equation \ref{dist}.

The magnitude distribution has a peak at V-band magnitudes between 20
and 23. Note that we have neglected extinction here.  To understand
the extent to which extinction is important, we make use of the model
in Bahcall \& Soneira (1980).  The spatial distribution of the CWDBs
is important here. If the CWDB distribution trace the galactic disk as
a thin disk, the extinction is significant: as shown with a dot-dashed
line in Fig. 2, with $z_0=200$ pc in equation \ref{dist} the extinction can be high as 10 magnitudes,
especially towards the galactic center where most CWDBs will be
located. If the vertical distribution is allowed to expand, such as to
a thick disk with $z_0=1.5$ kpc, then the extinction, on average, is about 3 magnitudes and the resulting
modification still allows for a  substantial number of binaries
brighter than 25th magnitude. Such a thick disk has been suggested to 
explain the presence of high-velocity white dwarfs (e.g., Reid et al. 2001).

While extinction becomes a significant problem for the optical
identification  of CWDBs, there is an additional effect resulting
from tidal heating in close binaries that can potentially
increase their luminosities (Iben et al. 1998). For our standard case,
with no extinction, we calculate the distribution of magnitudes with
luminosities corrected for tidal brightening. For simplicity, we
consider the maximum brightening given in Iben et al. (1998), again
assuming equal mass binaries, as a function of the orbital period. 
With a gain of $\sim$ 3 magnitudes in
the overall distribution, there is also a substantial increase in the
number of binaries brighter than 20th magnitude.  In general, tidal
heating is not likely to be uniform and could produce hot-spots, which
would in turn produce features in the light curve. Radiation from the
companion star would produce similar effects.

In practice, the optical follow-up studies of LISA error boxes can be
conducted as a two step process.  First, the gravitational waves
provide information related to the period and the phase, which is
adequate to determine when eclipses in the optical light curve
are expected. While the orbital periods are at the level of 500
seconds, for a quick identification, one can consider phased optical
observations to identify the source with known period and eclipse
phase. Once the source has been identified, more detailed light curves
could be taken, to determine the orbital parameters, sampled at a few
tens of seconds or less. Such a light curve would solidify the
agreement between LISA's period and phase with that of the optical
data and help secure the identification exactly.

Since the magnitude distribution is expected to peak at magnitudes
fainter than 25, detailed study may be limited to the tail of the
distribution at the bright end. In general, with four-meter class
telescopes, we expect that a detailed analysis of light curves would
be possible for a magnitude limit of $\sim$ 22 to 23 with signal-to-noise
ratios of order 10 in integration times of order 25
seconds. With phase matching, this signal-to-noise ratio can be improved.
Given the follow-up error boxes number $\sim$ 3000, a dedicated telescope 
is not required for this
purpose. Because the binary inclination and distance are determined in the
GW parameter extraction process, and the brightest systems are also
generally the closest (see Fig.~2),  candidate bright-end CWDBs can be
preferentially selected such that fewer boxes may need to be
followed up.  However, even non-detections would provide information on the presence
of tidal heating for the highest-frequency binaries.

An imaging instrument capable of rapid (10--100 sec resolution)
monitoring is needed. The field of view should be at the level of a square degree or more
to match the error box sizes. Searches with smaller field of view instruments would be
time-consuming. 
Visual magnitude limits of 25 or 26, in 25 second or so time intervals, are desirable, and
a brighter limit would cut down the number of detectable sources
substantially. It is unlikely that such an instrument currently
exists, but given the potential of \LISA for CWDB science, its
planning is warranted and can be considered for upcoming telescopes such as
the Large Synoptic Survey Telescope (LSST), which, by design, has the desirable field-of-view for
a project such as this.

Additional information on the binary could be extracted with
spectroscopic studies that involve radial velocity measurements. The
radial velocities will be substantial, around a few times 100 km
s$^{-1}$. Time-resolved spectroscopy of faint WDs, with notoriously
few suitable spectral lines, could prove difficult for all but the
most nearby systems. However, time-averaged spectra will be able to
provide information on WD temperatures, which can be used in
combination with distance and magnitude information, as well as
eclipse data, to put better constraints on the WD parameters.


To summarize, the Laser Interferometer Space Antenna (\emph{LISA}) is
expected to individually detect  between $\sim$ 3000 and 6000 Galactic
close white dwarf binaries (CWDBs) through their gravitational
radiation at frequencies above 3 mHz.  Most of these binaries will be
localized with \LISA data to an accuracy that ranges from tens of
arcminutes to a few degrees.  Due to the small binary separation, the
light curves of a few tenths of these resolved CWDBs are expected to
show eclipses. This will give a unique signature for identification in
follow-up optical studies of \LISA error boxes, because the orbital
period and phase will be known from \LISA data.  To optically identify
a substantial fraction of CWDBs that are crudely localized with \LISA,
one will require a rapid monitoring campaign capable of imaging a
square degree or more at intervals of tens of seconds, down to
magnitude levels between 20 to 25. While such a study can be carried out with
four or more meter-class telescopes down to magnitude levels of 25, no instrument currently exists to image a
wide-field of view of a square degree or more continuously with sufficient time resolution and we highly
encourage an investment in this aspect given the scientific potential
related to \LISA data.
\vspace{0.5cm}

{\it Acknowledgments:} 
We thank Gijs Nelemans and  Sterl Phinney for useful discussions.
This work is supported by the Sherman Fairchild foundation and DOE DE-FG
03-92-ER40701 (AC).

\end{document}